\documentclass[12pt]{article}
\usepackage[utf8]{inputenc}
\usepackage[bitstream-charter]{mathdesign}
\usepackage{graphicx}
\usepackage{url}
\usepackage{graphics}
\usepackage[dvipsnames]{xcolor}

\newcommand{\revision}[1]{{\color{Black}{#1}}}

\vspace{-5mm}
\title{Comparing high-dimensional neural recordings by aligning their low-dimensional latent representations}
\vspace{-1mm}
\author{Max Dabagia\footnote{School of Computer Science at the Georgia Institute of Technology.},~ Konrad P.~Kording\footnote{Depts of Bioengineering and Neuroscience at the University of Pennsylvania.},~  Eva L.~Dyer  \footnote{Department of Biomedical Engineering at the Georgia Institute of Technology. <evadyer@gatech.edu> }}

\date{}

\usepackage{natbib}
\usepackage{amsmath}
\usepackage{lineno}
\begin{document}

\maketitle
\vspace{-6mm}
\begin{abstract}
\vspace{-1mm}
\noindent 
Many questions in neuroscience involve understanding of the  responses of large populations of neurons. However,  when dealing with large-scale neural activity, interpretation becomes difficult, and comparisons between two animals, or across different time points becomes challenging.
One major challenge that we face in modern neuroscience is that of {\em correspondence}
e.g. we do not record the exact same neurons at the exact same times. Without some way to link  two or more datasets, comparing different collections of neural activity patterns becomes impossible. Here, we describe approaches for leveraging shared latent structure across neural recordings to tackle this correspondence challenge. We review algorithms that map two datasets into a shared space where they can be directly compared, and argue that alignment is key for comparing high-dimensional neural activities across times, subsets of neurons, and individuals.
\end{abstract}

\vspace{-3mm}
\section{Introduction}
\vspace{-2mm}
The study of brain function and cognition relies on  the measurement and interpretation of {\em changes} in the brain's activity. Whether between states of disease and health \citep{frere2018alzheimer}, across novice- and expert-level performance on a complex task \citep{sadtler2014neural}, or in wakefulness versus sleep \citep{hengen2016neuronal}, we need ways to {\em compare neural activity} patterns and draw conclusions about how two neural recordings differ. 

Historically, changes in neural activity patterns have been examined at the level of single cells, by characterizing how the firing rate (or tuning curves) of each cell shifts across conditions \citep{hubel1959receptive,saxena2019towards}
%\citep{hubel1959receptive,roxin2011distribution, chariker2016orientation, stringer2016inhibitory}
. 
%\eva{are these latter works only looking at single cell responses? lets make sure} 
However, we know that the brain consists of highly interconnected networks of neurons \citep{bullmore2009complex}, and thus the impact of different disease states or conditions is likely to have complex effects across an entire circuit. Without modeling these higher-order dependencies between neurons, it may be impossible to model the collective impact of a given perturbation or transformation of neural state \citep{golub2018learning,kaufman2014cortical,russo2020neural}. 

At first glance, it may be tempting to attempt to compare two recordings of neural activity by pairing off similar neurons across recordings and then comparing the changes in the activity across these pairs. However, in many (if not most) cases we may be technically unable to do so (as we can not generally choose which cells we record), or it may be fundamentally impossible to do so (the brain changes over time, and different individuals might not share neurons that are identical).  The alternative is to summarize each recording in a way which is \emph{invariant to the specific subset of neurons that are polled.}
Many existing approaches build invariance through the computation of metrics that capture essential properties of the pairwise interactions between neurons (like clustering or density of connections) \citep{soudry2015efficient, nonnenmacher2017extracting, brinkman2018predicting}, or extract a global property of the network such as criticality \citep{chialvo2010emergent,ma2019cortical}. While these approaches can be informative when we know the right metrics to compare across conditions, a data-driven strategy to identify couplings across multiple units would provide an even richer multi-dimensional picture of changes in neural circuits.

\begin{figure}[t!]
  \centering
  \includegraphics[width=0.7\textwidth]{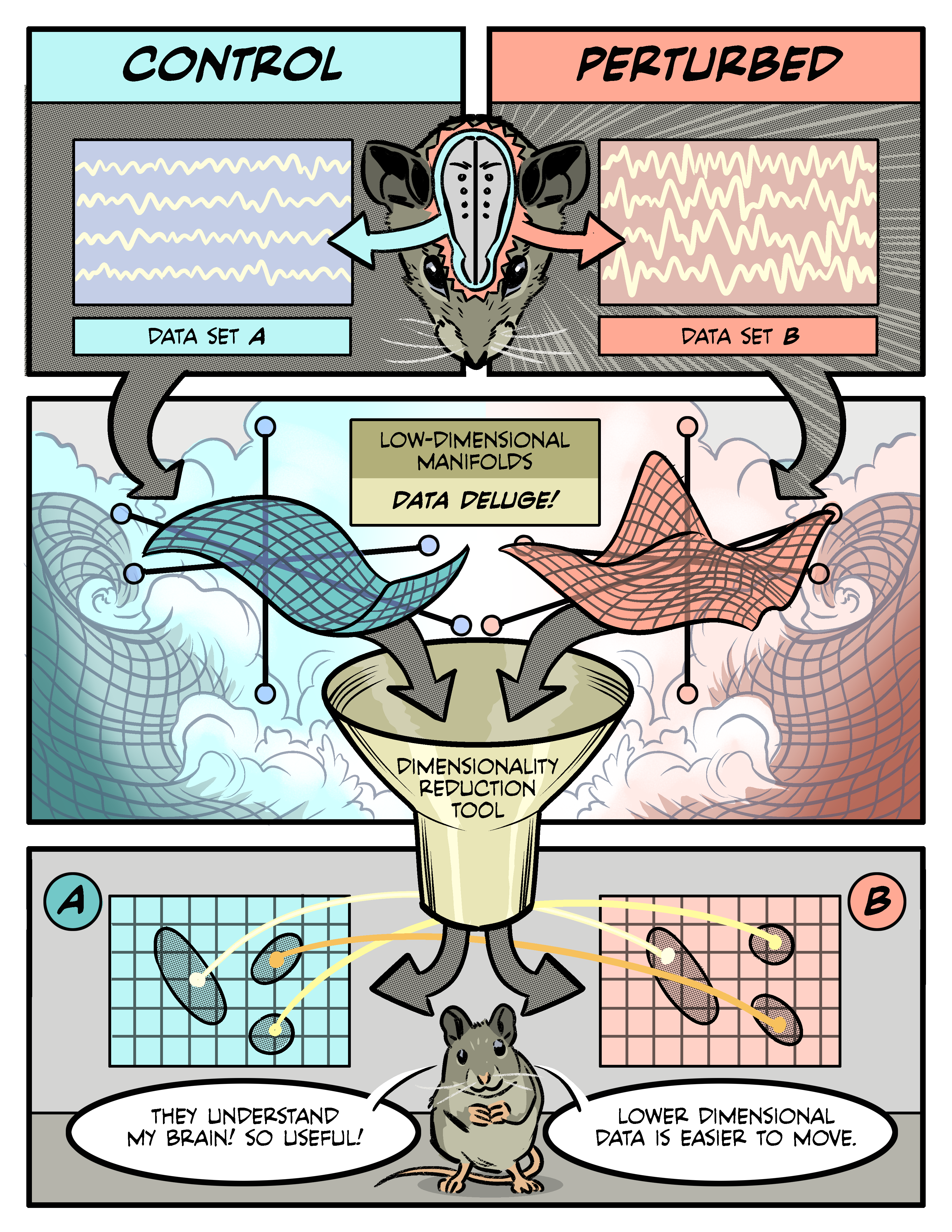}
  \caption{\footnotesize
  {\em Comparing neural activity patterns generated across two conditions through the alignment of their latent spaces.} In this example, we show the generation of two latent spaces (bottom) where datasets A and B share common structure but also have unique differences that can be detected.}
  \label{fig:fig0}
\end{figure}

One potential approach to tackle this correspondence challenge is to first find a representation of neural activity which is more abstract than the firing of individual units, and then compare different neural datasets through their representations (Figure~\ref{fig:fig0}). We can construct suitable representations by approximating the activity of each neuron as driven by a combination of a smaller number of \emph{latent factors} -- signals which describe coordinated activity across the population \citep{cunningham2014dimensionality}. 
Once each dataset has been fitted with a latent model, we hope that when the circuits are performing similar tasks, their latent models will agree with each other, allowing them to be compared. A substantial body of evidence shows that there are many settings in which latent factors are stable (or can be tracked) across subsets of neurons, across time, and even individuals  \citep{churchland2012neural, mante2013context, kaufman2014cortical, dyer2017cryptography, gallego2018cortical, pandarinath2018inferring,williamson2019bridging,gallego2020long,degenhart2020stabilization}, suggesting that these models may be a powerful tool for comparison. 
%The issue is that the mapping from latent factors to neural activity is subject to change. %While potentially invariant to , the set or ordering of latent factors is generally not uniquely defined, and so may differ superficially when calculated for several recordings separately. 

The roadmap for the rest of this article is as follows: In Section~\ref{sec:whyalign}, we discuss settings where latent space alignment is needed to effectively compare multi-dimensional neural activities across times, subsets of neurons, and individuals. After introducing the idea behind latent space alignment, in Section~\ref{sec:approaches}, we describe and compare different approaches for aligning neural datasets. In Section~\ref{sec:challenges}, we discuss  the challenges that must be overcome to compare recordings at larger scales and across more severe perturbations. Finally, in Section~\ref{sec:outlook}, we end with prospective analyses and discoveries that these approaches will make possible.
%We argue that alignment is key for comparing high-dimensional neural activities across times, subsets of neurons, and individuals.

\vspace{-3mm}
\section{Why we need alignment}
\label{sec:whyalign}
%In this section, we argue for why alignment is needed to begin to compare representations of neural activity patterns.

\subsubsection*{Forming a latent state space model}
To build a population-level view of neural activity, we start by representing the firing rates of all measured neurons as a point in a multi-dimensional state space, where each axis represents the firing rate of one of the recorded neurons (Figure~\ref{fig:fig2}a) \citep{paninski2010new}. Viewed this way, the dimension of this space is simply the number of neurons sampled in the recording. If every neuron operated completely independently this would be the end of the story, and moreover neuroscience would be in trouble, because understanding a neural circuit would require the characterization of every single neuron involved in it. Fortunately, there is \revision{a substantial body of} evidence that this is not the case 
\revision{\citep{mazor2005transient, mante2013context, kaufman2014cortical}}. Intuitively, neurons interact with each other through both excitation and inhibition, which naturally leads to correlation in their activity \citep{zohary1994correlated, abbott1999effect}. When many neurons are correlated with each other, it is reasonable to expect that their activities might be explainable by a smaller number of {\em latent} variables.

The simplest methods for building such a model identify each latent variable with a particular pattern of neural activity, such that that any observation can be approximated as a weighted sum of these factors \citep{tsodyks1999linking, sadtler2014neural, luczak2015packet}. Principal Components Analysis (PCA) is probably the most widely-used technique \citep{kaufman2014cortical, harvey2012choice, cowley2016stimulus}, which chooses latent variables which retain the most variance of the observations. There are many extensions of this approach that either relax PCA’s assumptions (e.g. that the latent variables must be orthogonal) \citep{child1990essentials}, or incorporate additional structure into the solution (e.g., adding a non-negativity constraint) \citep{lee1999learning, hyvarinen2013independent} to better model neural data. More sophisticated manifold learning strategies can be used to identify low-dimensional structures that may be nonlinearly embedded in neural activity, and ``flatten'' them to recover the latent factors \citep{stopfer2003intensity, ganmor2015thesaurus}.

The methods mentioned above assume that time points are independent and thus don't necessarily make use of temporal dynamics. Rather than modeling each point independently, we can instead model neural activity as being governed by a dynamical system to find a low-dimensional representation which preserves  coherent dynamics \citep{gao2017theory, kobak2019state, churchland2012neural, brunton2016extracting}. A state-of-the-art approach for learning nonlinear latent dynamics, called LFADS \citep{pandarinath2018inferring}, provides a deep learning solution to this problem that builds a realistic generative model of neural spiking using a sequential autoencoder. This approach can then be used to identify stable low-dimensional structure in neural activity across multiple individuals performing the same motor task. Incorporating temporal information can yield more stable latent factors, and crucially more interpretable ones.

\subsubsection*{Why can't latent spaces be compared directly?}
The critical assumption for using a latent factor model to compare two neural recordings is that their latent spaces provide an adequate representation of what is driving the circuit's activity. %While, in general, this is not possible to confirm, to evaluate the quality of representations, one often defines a decoding task that can be used to determine whether relevant task variables (i.e., input stimulus or motor output) can be predicted from the learned latent space. 
For the purposes of comparing two sets of neurons, we assume that if they encode similar information there should exists views of their latent spaces in which their activity looks similar. The objective is to find these views and put both datasets into a common reference frame.

\begin{figure}[t!]
  \centering
  \includegraphics[width=0.86\textwidth]{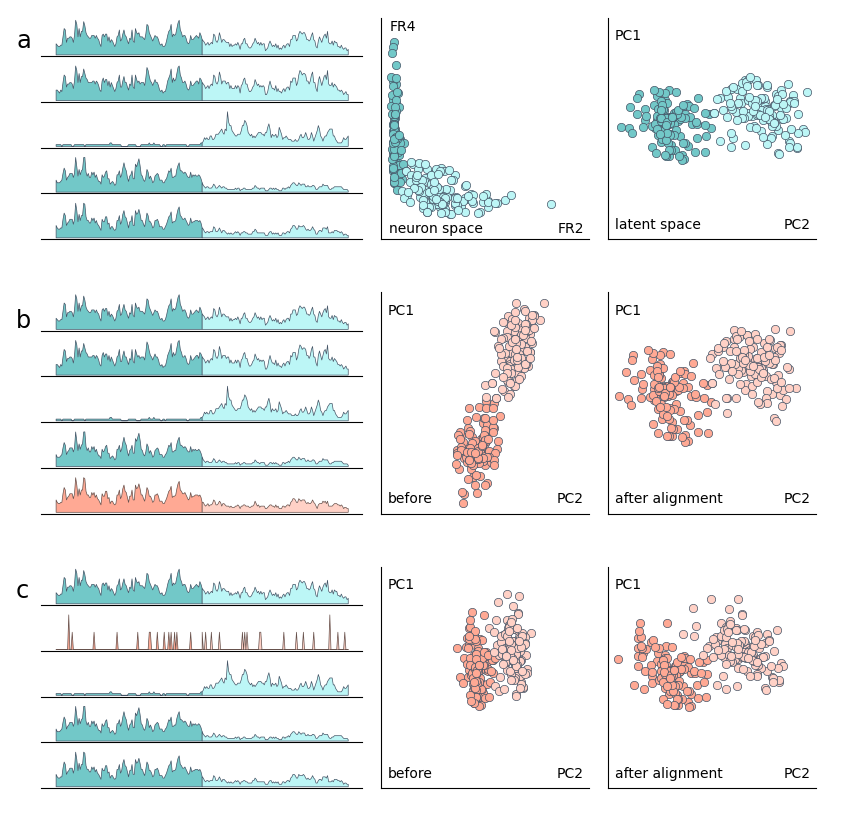}
  \caption{\footnotesize
  {\em Perturbations to neural activity can radically alter the latent space. } In (a), we show an example of unperturbed neural activities of 5 neurons, with the firing rates of two neurons displayed on the left and after the population is transformed into 2D with principal components analysis (PCA). In (b-c), we show examples of two different perturbations that are often observed in neural recordings. In (b), the baseline firing rate of the fifth neuron is increased by 1 spike per time interval (highlighted in pink). In (c), the activity of the second neuron is replaced with random noise, simulating its death or loss of contact with the electrode. In all cases the latent space displays the same structure, although the perturbed datasets have experienced an information-preserving transformation which must be reversed via alignment. To the right in (b) and (c), we show each perturbed latent space after alignment with Hierarchical Wasserstein Alignment (HiWA) \citep{lee2019hierarchical} to the match the distribution obtained in the first baseline dataset.
  }
  \label{fig:fig2}
\end{figure}

Unfortunately, even when they exist, finding views of two neural recordings that situate similar brain states in nearby parts of the latent space is typically not possible without further intervention or assumptions.  This is due to the fact that each dimensionality reduction technique relies on assumptions about how information is encoded to differentiate signal from noise and infer latent factors. Without a complete understanding of the brain's encoding strategies, these assumptions are at best approximations. For example, PCA associates latent factors with the patterns of activity which account for as much of the population's variance as possible. When the patterns of interest are roughly equally important, relatively insignificant changes in the activity of a few neurons can reorder these patterns, leading to latent models which encode the same information but must be transformed to match each other (Figure \ref{fig:fig2}). In more sophisticated methods, the ordering of factors is typically not unique, and comes down to particularities of the algorithm. Moreover, common factors may be obscured by additional neural activity, which is unrelated to the information shared across the recordings.

Adding to these challenges, in long-term recordings, instabilities arise from the difficulty of tracking particular neurons \citep{grill2009implanted, mccreery2010neuronal, fu2016stable} or matching them across time after sorting, and this can also introduce changes in which factors are dominant. In comparisons across individuals, or in acute recordings, direct comparisons across neurons is impossible. All of these challenges make direct comparisons across latent factors difficult and highlights the need for alignment: we must ensure that similar brain states in the two datasets are placed in the same part of the space.

\section{Finding correspondence across distinct neural recordings}
\label{sec:approaches}
In this section, we discuss different approaches for alignment. These approaches rely on different assumptions about the shared structure of the two datasets, or partial information about correspondences, either across neurons or across time.
%first class of methods, which we refer to as {\em manifold alignment} techniques, try to match the distribution or subspaces underlying two datasets. The second class of methods use {\em dynamics}, or different aspects of the temporal structure shared between two datasets to bring them into alignment. 

\subsubsection*{Approach 1. Using the global distribution of latent states for alignment}
If we assume that the latent spaces are comparable but they may have a different ordering of their latent factors, then this introduces the need for building in rotational invariance, or finding the best rotation of the latent factors that aligns the source and target data maximally (Figure \ref{fig:fig2}b-c, right).
One way to pose this problem is to find a mapping between \revision{the neural representations} which \revision{\emph{aligns} them -- that is, makes them maximally similar under some measure of similarity}. 
%\revision{A few different statistical metrics are commonly used, which frame a recording as a probability distribution over the latent space.} 
The space of mappings to consider should account for the invariances we require, and may be tailored to the individual problem.

Distribution Alignment Decoding (DAD) \citep{dyer2017cryptography}, uses density estimation to infer the distribution of neural activity in the latent space, and then, searches over the space of rotations to identify the transformation that best matches the two distributions based upon their Kullback-Liebler (KL) divergence. This idea has been applied to achieve nearly unsupervised neural decoding of movement, by aligning the low-dimensional projections of neural activity to a  distribution of movements without known correspondences between the two. Hierarchical Wasserstein Alignment (HiWA) \citep{lee2019hierarchical} improves upon this strategy by leveraging the tendency of neural circuits to constrain their low-dimensional activity to clusters or multiple low-dimensional subspaces. Combined with with a more geometric formulation of distance drawn from optimal transport \citep{villani2008optimal}, HiWA is able to more quickly and robustly recover the correct rotation to align latent spaces across neural recordings. 

A deep learning-driven solution to the alignment problem \citep{farshchian2018adversarial} draws inspiration from generative adversarial networks (GANs) \citep{goodfellow2020generative}, which have received some attention in the computational neuroscience community both for generative tasks like producing realistic neural activity \citep{arakaki2017capturing, molano2018synthesizing}, or to aid in downstream tasks like decoding visual stimuli \citep{st2018generative}.  \revision{Here, the idea is to start with a latent model for one dataset from which one can reconstruct neural activity effectively, and then transform new datasets to match by minimizing their reconstruction loss using the same latent model.} This approach is promising for brain-machine interfaces (BMIs), where the priority is correctly decoding target variables and less about interpreting the \revision{differences in the learned embeddings}. On the other hand, when using latent representations to study a neural circuit, it might be desirable for distinctive new features of the shifted distribution to be preserved, but a sufficiently powerful alignment network which has ``overfit'' to the target dataset will attenuate them in favor of matching the original dataset as closely as possible. 

\subsubsection*{Approach 2. Using partial neuron correspondences for alignment}
This same idea of rotation-invariant manifold alignment has recently been demonstrated in an online BMI setting \citep{degenhart2020stabilization}. This method assumes a sufficient number of neurons are captured across subsequent recordings, so that only a few coordinate axes of the neural manifold need to be adjusted. The stable units are estimated to be those which have approximately the same relationship with the latent factors, while the perturbed units are remapped using a constrained transformation. Although less general than distribution alignment because it assumes the underlying transformation to be a permutation of a few of the coordinate axes and that the identities of some neurons are known, this assumption also decreases the computational requirements, making this solution more practical in a realistic online BMI setting. When used to incrementally update the decoder for intra-cortical multi-electrode arrays \citep{chestek2011long,vaidya2014ultra} (and their  successors), this strategy is likely to be a crucial building block for stable long-term BMIs, especially when used in conjunction with covariance inference techniques used in calcium imaging \citep{turaga2013inferring, nonnenmacher2017extracting} to stitch together recordings across large populations to fill in manifolds.

\vspace{-2mm}
\subsubsection*{Approach 3. Using temporal correspondence for alignment} 

The methods described in the last section assume that we have common global distributional or subspace structure between datasets to align them. However, there are many cases where this assumption may be insufficient or incorrect. Instead, when we have some amount of temporal correspondence between datasets, either through simultaneous recordings \citep{sponberg2015dual}, or by using the temporal structure which arises from multiple trials of the same task \citep{gallego2020long}, then we can use information to embed multiple datasets jointly (Figure \ref{fig:fig3}). 
%When temporal correspondence is possible, pairing off data samples and using it to learn latent representations can result in latent models which preserve more of the relevant information than an independent method would. 
A straightforward approach for doing so 
%that has withstood the test of time 
is canonical correlation analysis (CCA) \citep{hotelling1936relations, fujiwara2013modular, dmochowski2018extracting, gallego2020long}, and more sophisticated nonlinear extensions \citep{lai2000kernel, huang2006kernel, andrew2013deep}. This technique is a generalization of PCA, where rather than choosing a single factor (a pattern of neural activity) at a time which maximizes variance, it chooses a factor for \emph{each dataset} so that the covariance between these components is maximized. CCA reduces to PCA when the datasets being compared are the same. The problem is that CCA requires exact correspondence between moments in time across datasets, which, unless they were made simultaneously, is difficult to satisfy. Nonetheless, due to its simplicity and efficiency, CCA has been widely used in neuroscience, even when this key assumption is only loosely satisfied (such as when each recording is a trial of a precisely timed task) \citep{gallego2018cortical,gallego2020long}.

\vspace{-2mm}
\subsubsection*{Approach 4. Using common dynamics for alignment} 
Temporal information can be used more generally by finding a latent space in which the dynamics of each dataset are similar \citep{pandarinath2018latent}. For instance, in LFADS \citep{pandarinath2018inferring}, the latent representations of dynamics from many recording sessions can be matched to each other by learning a shared latent space for all sessions jointly in which temporal dynamics are consistent. This method develops the mappings between each recording session (dataset) and the latent space by first finding the trial-averaged responses for each unit and then learning a weight matrix (input map) to match these trial-averaged responses. This means that given training data consisting of the direction (or target) of the reach, all of the trajectories in the same direction can be utilized to improve the quality of the aligned latent representations, which alleviates the need to maintain contact with the same neurons across recordings. In principle, however, this approach could also be fully unsupervised, or use partial information about clustering of behaviors (like HiWA) to find alignment of dynamics.  
%A common latent space model is learned across all of the measured units over all days and used to decode on any one of the days. 
By leveraging shared task structure across datasets, methods like LFADS show that it is possible to stitch together dynamics from different populations of neurons and decode intended movement variables.

\begin{figure}[t!]
  \centering
  \includegraphics[width=.86\textwidth]{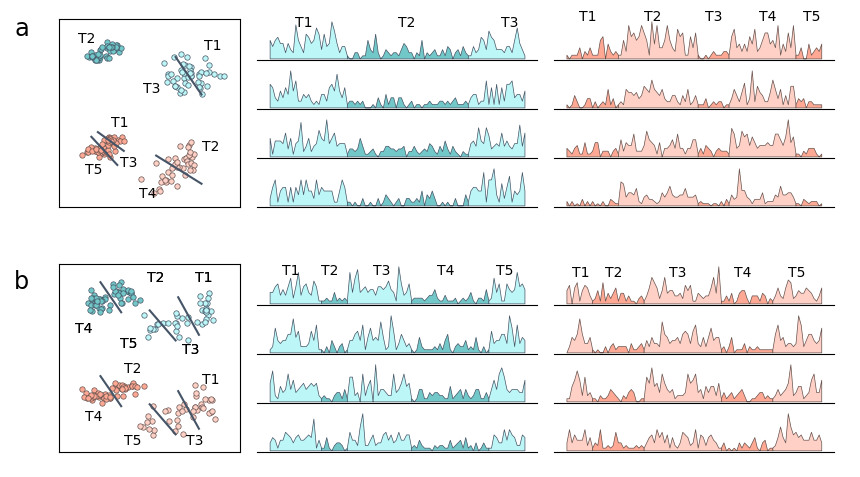}
  \caption{\footnotesize
  {\em Alignment Paradigms.} In both (a-b), we compare distribution alignment with strategies that leverage temporal structure and dynamics to align data into a common latent space. In (a), we show an example where in the baseline condition (blue), neural activity traverses across three subspaces in the latent space (T1, T2, T3); in the second condition (pink), the individual makes a very different traversal through the latent space, jumping between different subspaces and not adhering to the same dynamics as in the first case. However, the overall global structure of the distribution is similar in both cases and thus manifold alignment techniques can be applied successfully. In (b), we show an example where the temporal structure is preserved between both datasets by having a consistent sequence of transitions between activity modes. In this case, we do not need to assume the same overall global structure, but that dynamics are consistent.
  %This is accomplished through joint dimensionality reduction and alignment, which finds a latent space in which the temporal structure matches. 
  }
  \label{fig:fig3}
\end{figure}

\revision{
\vspace{-2mm}
\subsubsection*{Comparison of various approaches} 
Given how challenging it can be to read out the same information when it has been encoded in multiple unknown complex dynamical systems, each of the methods discussed previously relies on a different set of assumptions to make the problem tractable. Distribution alignment exploits  similarities between the data embedded in their latent spaces to bring them into agreement; this typically requires that there are sufficient similarities in the behavior over time and that a sufficient number of samples are available to decipher this structure (Figure~\ref{fig:fig3}a). This places a hard requirement on the latent structure; if the similarities are absent, too few, or too imprecise, the alignment problem becomes ill-posed. However, when this latent structure exists, these methods can align recordings separated by large amounts of time or even from different individuals. In contrast to this more global approach, approaches that leverage dynamics in the progression of neural states (Figure~\ref{fig:fig3}b), may take a more local view of modeling neural activity patterns which places constraints on how two datasets must evolve over time. When additional information is available which allows datasets to be matched to each other, such as temporal or partial neuron correspondence, alignment techniques can leverage this information to succeed in cases too ambiguous for manifold alignment.}

\vspace{-2mm}
\section{Challenges}
\vspace{-2mm}
\label{sec:challenges}
An essential tool for comparing neural datasets is extracting latent representations that preserve key structures within them. However, using dimensionality reduction approaches to study neural population dynamics can be daunting because they are typically {\em unsupervised} and the latent factors learned by these methods can be hard to interpret. Even with supervised labels or training data (like movement kinematics, sleep  or behavioral state), it can be hard to know when a latent space model is sufficient to describe the structure of the data.

Furthermore, it is unclear when latent variables should be considered  intrinsic to the functionality of a neural population, or when they are just a concise description of activity. \revision{This is an instance of the classic statistical conundrum of differentiating causation versus correlation. Even between neurons with correlated activity, it is practically impossible to determine whether the correlation arises because one neuron drives the activity of another, or if there is another \emph{hidden} neuron driving both of them \citep{mehler2018lure}.}
For example, it is not understood whether the ability to decode movement patterns from activity in motor cortex is because neurons have been found which are designed to encode movement, or because their activity is highly correlated with movement as a side-effect of their true functional roles. Most dimensionality reduction techniques are agnostic to this distinction, \revision{yet hopefully we can distill neural activity down to a space that reveals its true underlying causes \citep{lillicrap2019does}. Moving forward, careful design of experiments that can modify neural response properties, either through either direct manipulation \citep{gradinaru2010molecular,roth2016dreadds} or new learning paradigms \citep{shenoy2021measurement}, are needed to better interpret neural representations, and understand how circuit and systems-level mechanisms modulate or change the distribution of the neural state space.}

% partial matching 
Most of the alignment methods covered in the previous section assume that there is observable low-dimensional structure that persists across multiple recordings. However, there are many settings where changes in underlying brain states may occur due to sources of modulation like attention and engagement \citep{mitchell2009spatial, cohen2009attention, zhang2011object,mcadams1999effects}, or shift as units die or electrodes move \citep{polikov2005response, grill2009implanted, mccreery2010neuronal}. 
%\revision{Latent space alignment provides some ways to overcome certain types of transformations in data, particularly those due to recording instability.
In more challenging settings where common structure may not be easily ascertained, some supervised data or labels could be provided to the methods to help ground alignment when common structure between latent spaces is not sufficiently similar to find a map between them. More transient modulations are harder to overcome, but methods developed for dealing with them for supervised decoding might provide the starting point for a way forward \citep{fan2014intention, sederberg2019state, whiteway2019characterizing}. Additionally, methods for partial domain adaptation \citep{cao2018partial, cao2019learning, lin2020making} may be utilized to facilitate matching in these conditions.  Methods that can account for realistic shifts in neural data will allow for comparisons, while also modeling state-dependent changes that may modulate the neural distribution \citep{feulner2021neural}.

As the algorithms to perform these types of analyses grow more sophisticated and specialized, it will be important to apply them to neural datasets that are generated across sufficiently diverse conditions. After all, carefully obtained datasets of this nature are the only way of knowing if the approaches we use actually work in practice. In machine learning, common datasets used across the community have often spurred progress, from simple benchmarks like MNIST \citep{lecun1998gradient} to more substantial datasets like the ImageNet corpus \citep{russakovsky2015imagenet}.
In neuroscience, open-access brain imaging datasets and tools like those provided in the Allen Institute’s optophysiology-based Brain Observatory \citep{de2020large}, and their more recent high-density survey with Neuropixel probes \citep{siegle2019survey}, could become benchmarks by which to compare methods, as well as fertile ground for more in-depth studies. Moving forward, widespread data sharing like this will allow for more reproducible results and pave the way for more powerful and generalizable algorithms.

\vspace{-2mm}
\section{Outlook}
\vspace{-2mm}
\label{sec:outlook}
In current brain machine interfaces (BMIs), decoders rely on stable mappings from neural activity to the variables of interest (e.g., position, speed, or direction of intended movement),  which often degrade over time \citep{pandarinath2018latent}. Thus with the ability to directly compare latent representations and align them, it will be possible to correct for changes in neural distributions to develop more robust neural decoding interfaces. These controllers might be able to adapt to shifts in neural activity due to cognitive state without collection of new training data, \revision{which is crucial for a controller which will be chronically implanted.}
We emphasize that latent space alignment should not be considered only a convenience for analysis (and label transfer), but also a tool for improving the weaknesses of current BMIs and neural prosthetics. 

While many of the approaches we discussed use linearity assumptions to simplify the challenge of representation alignment, the computations and associated latent spaces present within both small circuits and across larger neural populations are likely to lie in spaces that linear models cannot appropriately capture. In these cases, more complex models like unions of subspaces and manifolds \citep{elhamifar2011sparse,elhamifar2013sparse,dyer2013greedy} and statistical tests based upon these models may be used to compare more complex low-dimensional structures.  By coupling multi-manifold models with matrix completion methods \citep{bishop2014deterministic, liu2017new}, it could also be possible to stitch together manifolds from many brain areas and start to fill in the gaps across recordings that are collected from different sites of contact in the brain across non-overlapping populations of neurons. Thus, through combining the techniques described in this article with more complex latent space models, we may be able to compare whole brain activity patterns, even when individual recordings provide only a small view of the bigger picture. 

While understanding of the impact of neurological and neurodegenerative diseases has advanced tremendously in recent years \citep{frere2018alzheimer,wingo2021integrating}, a description of the disruptions they cause to the functionality of neural circuits has not yet been possible. It is highly likely that when such a description is made, it will be through the study of shifts in the activity of many neurons. The alignment strategies outlined here, and more sophisticated methods built on them, will provide new avenues to compare the latent spaces that the brain occupies in health and provide signatures of decline as it progresses into disease. 

\vspace{-2mm}
\section*{Code availability}
\vspace{-2mm}
% A notebook to create the synthetic example in Figure~\ref{fig:fig2} and align the baseline and perturbed conditions with HiWA \citep{lee2019hierarchical} are provided at: \url{nerdslab.github.io/neuralign/}. 
Demonstrations of the alignment methods DAD \citep{dyer2017cryptography} and HiWA \citep{lee2019hierarchical} and implementations in Python and MATLAB are available at: \url{nerdslab.github.io/neuralign/}

\vspace{-2mm}
\section*{Acknowledgements}
\vspace{-2mm}
We thank Benjamin Dewey (\url{https://benjamindewey.tumblr.com/}) for creating the awesome illustration in Figure 1, and Chethan Pandarinath and Mehdi Azabou for helpful feedback on this manuscript. This work was supported by NIH-1R01EB029852, as well as generous gifts from the Sloan Foundation and McKnight Foundation.

\bibliographystyle{authordate1}
\bibliography{references}
\end{document}